\documentclass[%
 preprint,
 amsmath,amssymb,
 aps,
]{revtex4-1}

\usepackage{graphicx}
\usepackage{dcolumn}
\usepackage{bm}

\usepackage{lipsum}
\usepackage{latexsym}
\usepackage{amsfonts}
\usepackage{graphicx}
\usepackage{epsfig}
\usepackage{amsmath, amsthm, amssymb}

\begin{document}

\title{Molecular Scale Hydrophobicity in Varying Degree of Planar Hydrophobic Nanoconfinement}

\author{Sudip Nepal and Pradeep Kumar}
\email{pradeepk@uark.edu}
\affiliation{ Department of Physics \\ University of Arkansas Fayetteville AR 72701} 
\date{\today}

\begin{abstract}

\noindent We have studied the molecular scale hydrophobicity of an apolar
solute, argon, confined between hydrophobic planar surfaces with
different confinement widths. Specifically, we find that the
hydrophobicity exhibits a non-monotonic behavior with confinement
width. While hydrophobicity is usually large compared to bulk value,
we find a narrow range of confinement width where the hydrophobicity
displays similar values as in bulk water. Furthermore, we develop a
simple model taking into account the entropic changes in nanoconfined
geometry, which enables us to calculate potential of mean force
between solutes as the conditions change from bulk to different
degrees of planar nanoconfinement. Our results are important in
understanding nanoconfinement induced stability of apolar polymers,
solubility of gases, and may help design better systems for Enhanced
Oil Recovery.
\end{abstract}

\pacs{Valid PACS appear here}
\maketitle

\noindent
\section{Introduction}
\bigskip

Confined systems are prevalent in nature. Examples of confinements
include intracellular water and biomolecules to gases and liquids
trapped in porous rock
sediments~\cite{Tang2006,Hartls2011,Wu2013}. One can expect that when
confinement length scales approach the inherent characteristic
molecular length scales, such as various correlation lengths, the
properties of a system may deviate from bulk~\cite{Kumar2005}.

It is widely believed that besides the interaction of solvent with
solutes, thermodynamic states and dynamics of solvent, water in case
of aqueous systems, are important in dictating the behavior of
solubility of solutes. Nanoconfinement induced changes in the
properties of water is quite extensively studied in recent
times~\cite{Koga2000,Koga2001,Truskett2001,Zangi2003,Gallo07,Johnston2010,Han2010,Chen2006,Liu04,Nicolas2007,Nicolas2009,Mallamace2011}. It
has been shown that when water is confined to length scales of few
molecular diameters, water exhibit distinct structural and dynamical
changes as compared to bulk water~\cite{Raviv2001}. Koga et. al. have
studied liquid-solid phase transitions in water confined between
hydrophobic surfaces and have shown that for appropriate dimensions of
the flat hydrophobic confinements, water can undergo a first-order
phase transition from a bilayer liquid to a bilayer hexagonal
ice~\cite{Koga2000,Koga2001}. Further studies of water in
nanoconfinements have suggested a range of exotic structural phases
such as n-fold crystal (n=2,4,6, and 12) in bilayer
water~\cite{Johnston2010}. Bai et. al. have found a variety of
polymorphic and polyamorphic transitions in bilayer water confined
between hydrophobic surfaces~\cite{Bai2012}. Recently, Han
et. al. have studied liquid-solid phase transitions in bilayer water
confined between planar hydrophobic surfaces~\cite{Han2010}. They find
that nature of liquid-solid phase transition changes from a
first-order transition at low densities to a continuous liquid-solid
transition at high densities. Besides the structural changes in
nanoconfinement, it has been shown that the physical property such as
temperature of maximum density (TMD) of water shifts to lower
temperature and lower density in the temperature-density plane as
compared to bulk water~\cite{Kumar2005}. Gallo et. al. have studied
the effect of nanoconfinement on the dynamics and thermodynamics of
water and found that the water confined in nanopores exhibits
different dynamical regimes depending on the distance of water from
the surface~\cite{Gallo07}. The effects of morphology and charges of
confining surfaces have been explored in
Ref.~\cite{Varma2010,Nicolas2007}.

Aqueous solubility and hydrophobicity of solutes is an active area of
research due to its implication in a large variety of physical
phenomena including protein stability and
folding~\cite{Wiggins1997,Chandler1999,Asbaugh2002,Dill2005,Garde2006,Buldyrev07,Zhou2002},
hydrate formation~\cite{Walsh2009}, and enhanced oil
recovery~\cite{Wu2013}. Studies of solubility of hydrocarbon solutes
in water suggests that they exhibit anomalous solubility in
water~\cite{Battino1982,Battino1986,Battino1987}. For example methane
solubility in water shows a minimum around $320$~K and increases on
both sides of this temperature. Low temperature increase of solubility
of apolar solutes is implicated in cold unfolding of proteins and
apolar polymers~\cite{Buldyrev07}. In more recent works, Mallamace
et. al. have explored the influence of water on protein properties
including protein folding and dynamic transition in proteins at low
temperatures~\cite{Mallamace2011,Mallamace2014}. Recent computational
studies have investigated the effect of nanoconfinement on the phase
behavior of oil-water mixture in an effort to improve the existing
enhanced oil recovery technologies~\cite{Wu2013}. Studies of
solubility and ordered phase formation of gases in solid ice phases
have attracted wide attention due to its importance in hydrocarbon
processing and sustainable energy
production~\cite{Sloan2008,Sloan2003,Walsh2009,Debenedetti2009}.

Although a large body of literature exists on the solubility and
hydrophobicity of apolar solutes in bulk water, the effect of
nanoconfinement on the solubility and hydrophobicity remains poorly
understood. Since both thermodynamic and dynamic properties of water
may vary in nanoconfinement, one may expect that the hydrophobicity
may also be different from the bulk . Water in biological systems is
under the conditions which differ greatly from the bulk hence the bulk
behavior of water may not be relevant under those
conditions. Therefore, a good understanding of the behavior of
hydrophobicity of any molecule would only be possible if a detailed
study of the these conditions are explored.

In this article, we study molecular scale hydrophobicity of an apolar
solute, argon, in planar hydrophobic nanoconfinement of varying length
scales. In {\bf Method} section, we describe the simulation methods,
in {\bf Results} section, we discuss the results and finally we
conclude with {\bf Summary and Discussion} section.

\section{System and Method}

We performed molecular dynamics simulations to calculate potential of
mean force (PMF)~\cite{McQuarrie} between argon atoms dissolved in
TIP3P (transferable interaction potential three
points)~\cite{JorgensenXX,jorgensen2} water-like solvent confined
between two structured planar hydrophobic surfaces separated by a
distance $L_{\rm z}$. The atoms on the surfaces were arranged in a
hexagonal closed packing with a distance $\sigma_S=0.339$~nm and they
do not interact with each other. The positions of the surface atoms
were restrained to their respective mean positions by a harmonic
potential with a force constant $10^4$~kJ/mol. In order to mimic
hydrophobic surfaces, carbon atoms were chosen to be surface
atoms. The interaction between the surface atoms and the oxygen of the
water molecule is modeled using $6-12$ Lennard-Jones (LJ) potential
$U(r)$
\begin{equation}
U(r) = 4\epsilon[(\frac{\sigma}{r_{SO}})^{12} -
  (\frac{\sigma}{r_{SO}})^{6}]
\end{equation}
where $r_{SO}$ is the distance between the surface atom and oxygen of
water molecule, and $\sigma = (\sigma_{\rm S}+\sigma_{\rm OW})/2$ and
$\epsilon = \sqrt{\epsilon_{\rm S}\epsilon_{\rm OW}}$. $\sigma_{\rm
  OW}$ and $\epsilon_{\rm OW}$ are the parameters of the Lennard-Jones
interaction between oxygen atoms of water
molecules~\cite{JorgensenXX,jorgensen2}.

\begin{figure}
\begin{center}
\includegraphics[width=8cm]{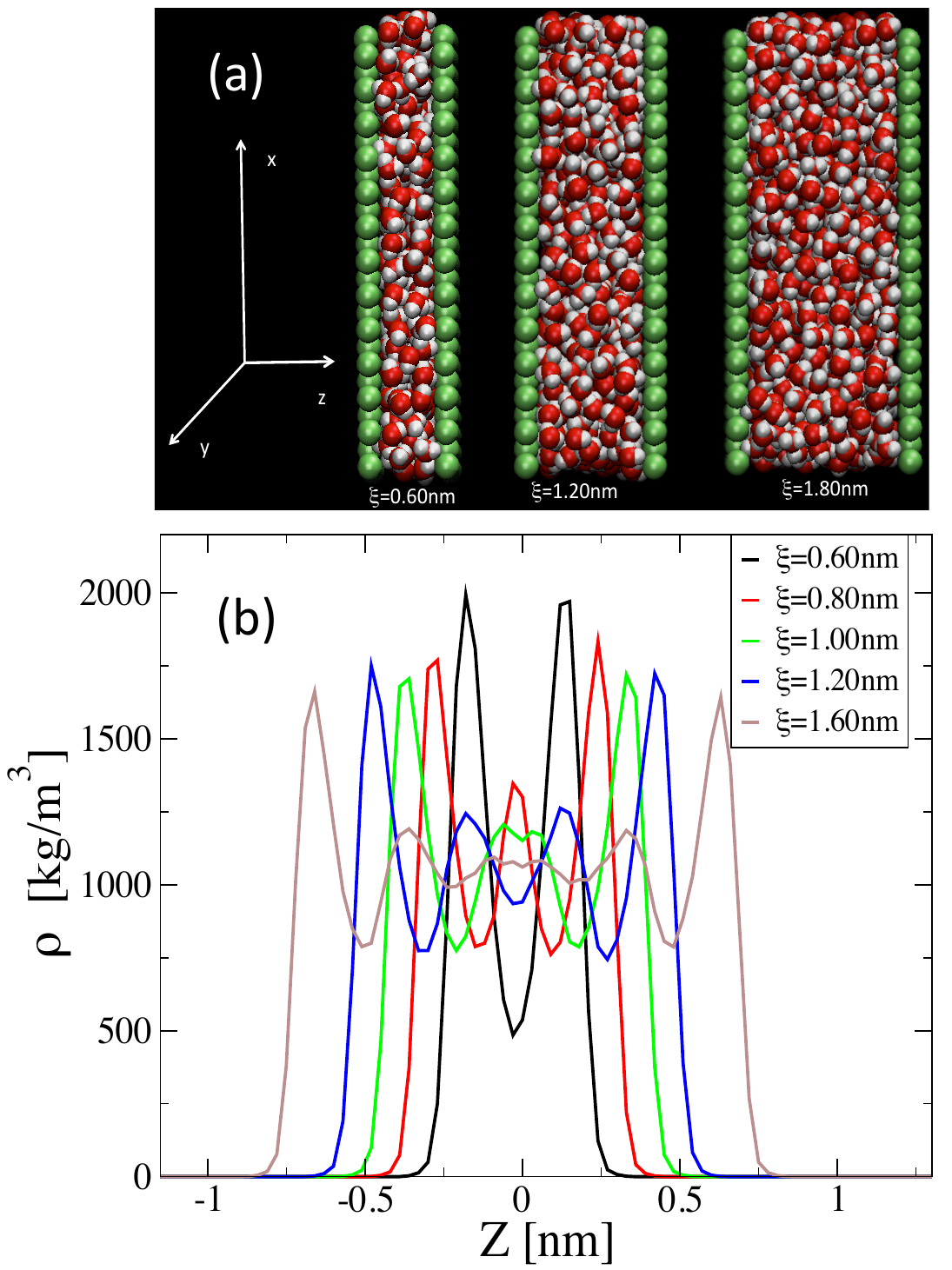}
\end{center}
\caption{(color online) (a) Schematic of nonoconfined system. (b)
  Potential of mean force (PMF) between two argon atoms as a function
  of distance $r$ between argon atoms for different confinement width
  $\xi$.}
\label{fig:fig1}
\end{figure}

We performed simulations of confined water and solutes between
hydrophobic surfaces in NVT-ensemble with effective density of the
system, $\rho=1.00$~g/cm$^3$. The effective density of the system is
obtained by calculating the effective confinement width available for
water molecules for a given center-to-center distance, $L_{\rm z}$,
between the surface atoms. The effective confinement width, $\xi$,
available to water molecules is calculated as
\begin{equation}
\xi=Lz-\frac{(\sigma_{\rm OW}+\sigma_{SO})}{2}
\end{equation}
The equations of motion were integrated with a time step of $0.001$~ps
and velocity rescaling was used to attain constant temperature and
Berendsen barostat for constant pressure in Gromacs
4.5~\cite{Gromacs}. Periodic boundary conditions were applied in
XY-directions. We use constrained molecular dynamics method for the
calculation of PMF for seven different effective confinement widths
$\xi=0.6,0.8,1.0,1.2,1.4,1.6$, and $1.8$~nm. Images of some of the
representative configurations studied here are shown in
Fig.~\ref{fig:fig1}(a).

After the system was equilibrated for $1$~ns at $T=300$~K for
respective confinements, we constrain the argon atoms at fixed values
of distances with a harmonic potential with a spring constant
$k_{c}=1000$~kJ/mol. After the equilibration step, we ran the
simulations for additional $1$~ns for each constraining distances $d$
between $0.26$ and $1.0$~nm at an interval of $0.02$~nm. Potential of
mean force was corrected for volume entropy at T$=300$~K. For the range of
confinement widths studied here, we find strong layering of water near
the surfaces as suggested by the transverse density profile of water
along the confinement direction $z$ (see Figure~\ref{fig:fig1}(b)).

\begin{figure}
\begin{center}
\includegraphics[width=9cm]{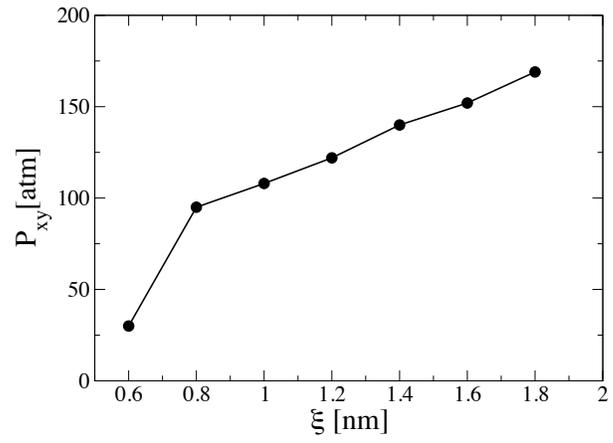}
\end{center}
\caption{Lateral pressure $P_{\rm xy}$ as a function of effective
  confinement width $\xi$ for the fixed density $\rho={\rm 1.0
    g/cm^3}$. $P_{\rm xy}$ decreases monotonically with confinement
  width.}
\label{fig:fig2}
\end{figure}

While the effective density of the system remains the
same, the lateral pressure along the periodic directions, $P_{\rm
  xy}$, varies for different confinement widths as shown in
Fig.~\ref{fig:fig2}. $P_{\rm xy}$ monotonically increases with
effective confinement width $\xi$.
\section{Results}
\section*{Potential of mean force and second Virial coefficient}

To quantify hydrophobicity, we first calculate PMF, $w(r)$, for
different confinement widths using umbrella
sampling~\cite{Eerden1989}. To use umbrella sampling, the distance
between the argon atoms is chosen as the reaction co-ordinate. Within
this scheme, a constraining potential is added to the Hamiltonian. The
modified Hamiltonian of the system with the constraining potential
with a perturbation parameter $\lambda$ can be written as
\begin{equation}
H(\lambda) = H_{0}+\frac{1}{2}k_c(r-r_{0}(\lambda))^2 
\end{equation}
where $k_c$ is the spring constant of the constraining potential and
$r_{0}$ is the constraining distance. The free energy change $\Delta
G$ between two arbitrary points along the reaction coordinate is given
by
\begin{equation}
\Delta G = \int{(\frac{\partial
    G}{\partial\lambda})_{\lambda}d\lambda}=\int{(\frac{\partial
    H}{\partial\lambda})_{\lambda}d\lambda}=\int{-\left
  <k_c(r-r_0)\right >_{r_0} dr_0}
\end{equation}
Hence the free energy difference between two states along the reaction
coordinate is integral of the mean force between the state points. The
PMF, $w(r)$, for a given state point along the reaction coordinate with
respect to a reference state is
\begin{equation}
w(r) = -\int_{r_{\rm ref}}^{r}{<F(r_0')>dr_0'}+2k_BT{\rm ln}(r)+C.
\end{equation}
where, $<F(r_0')>$ is the mean force for a given constraining distance
$r_0'$, $r_{\rm ref}$ is the position along the reaction coordinate for the
reference state, and the second term in the above expression is the
volume entropy correction. We choose the reference state to be
$r_{\rm ref}=1$~nm where $w(r_{\rm ref})=0$.

\begin{figure}
\begin{center}
\includegraphics[width=8cm]{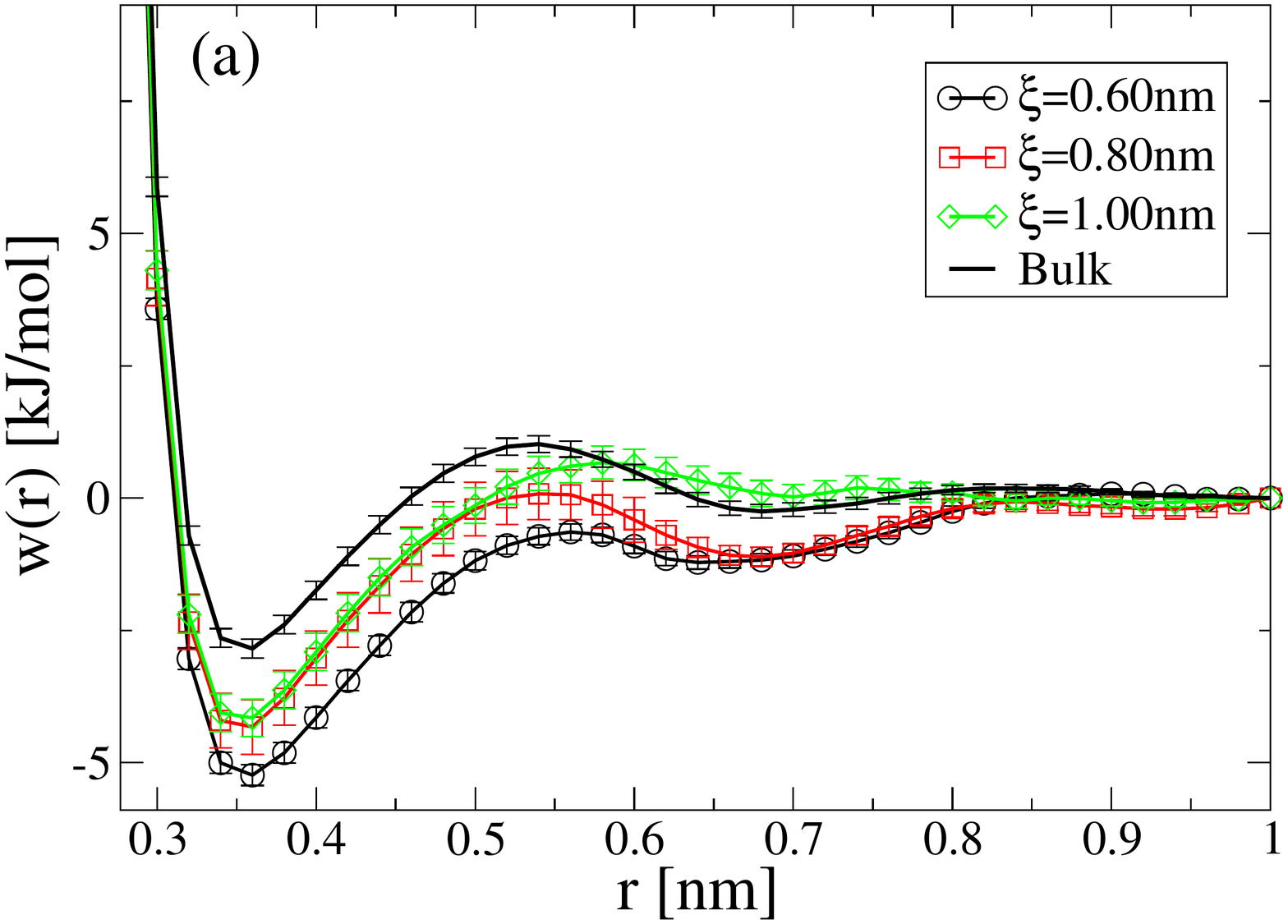}
\includegraphics[width=8cm]{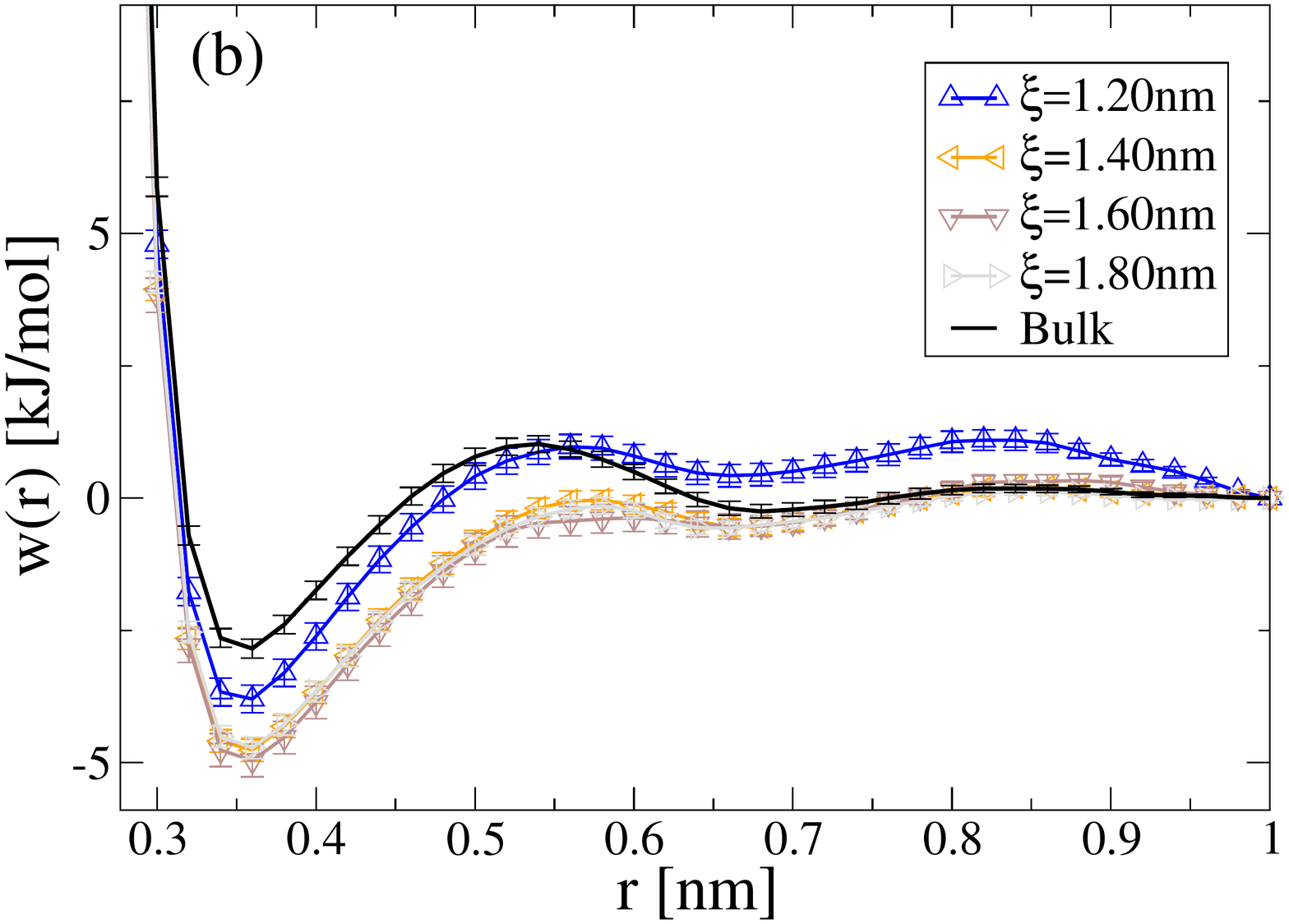}
\includegraphics[width=8cm]{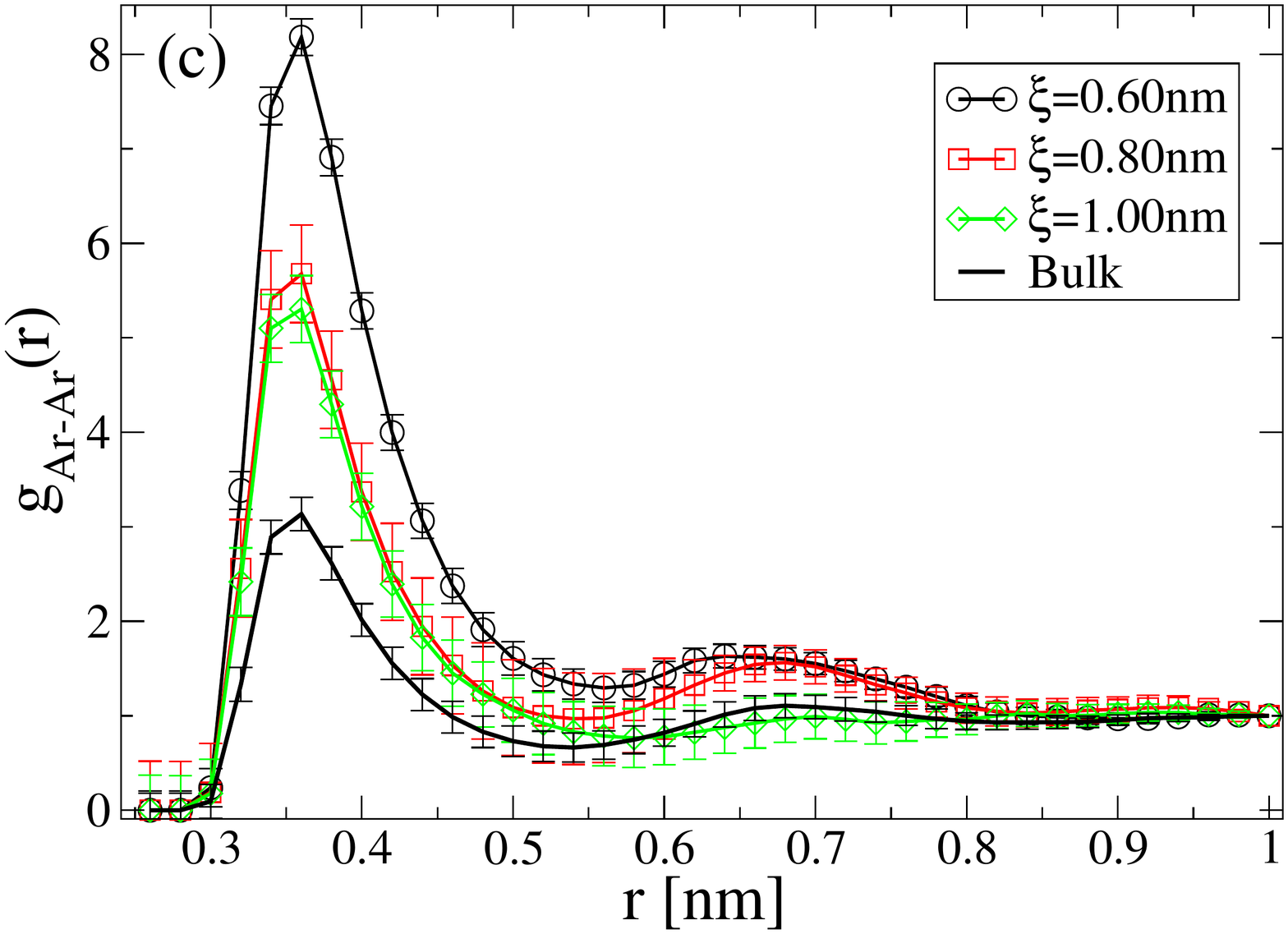}
\includegraphics[width=8cm]{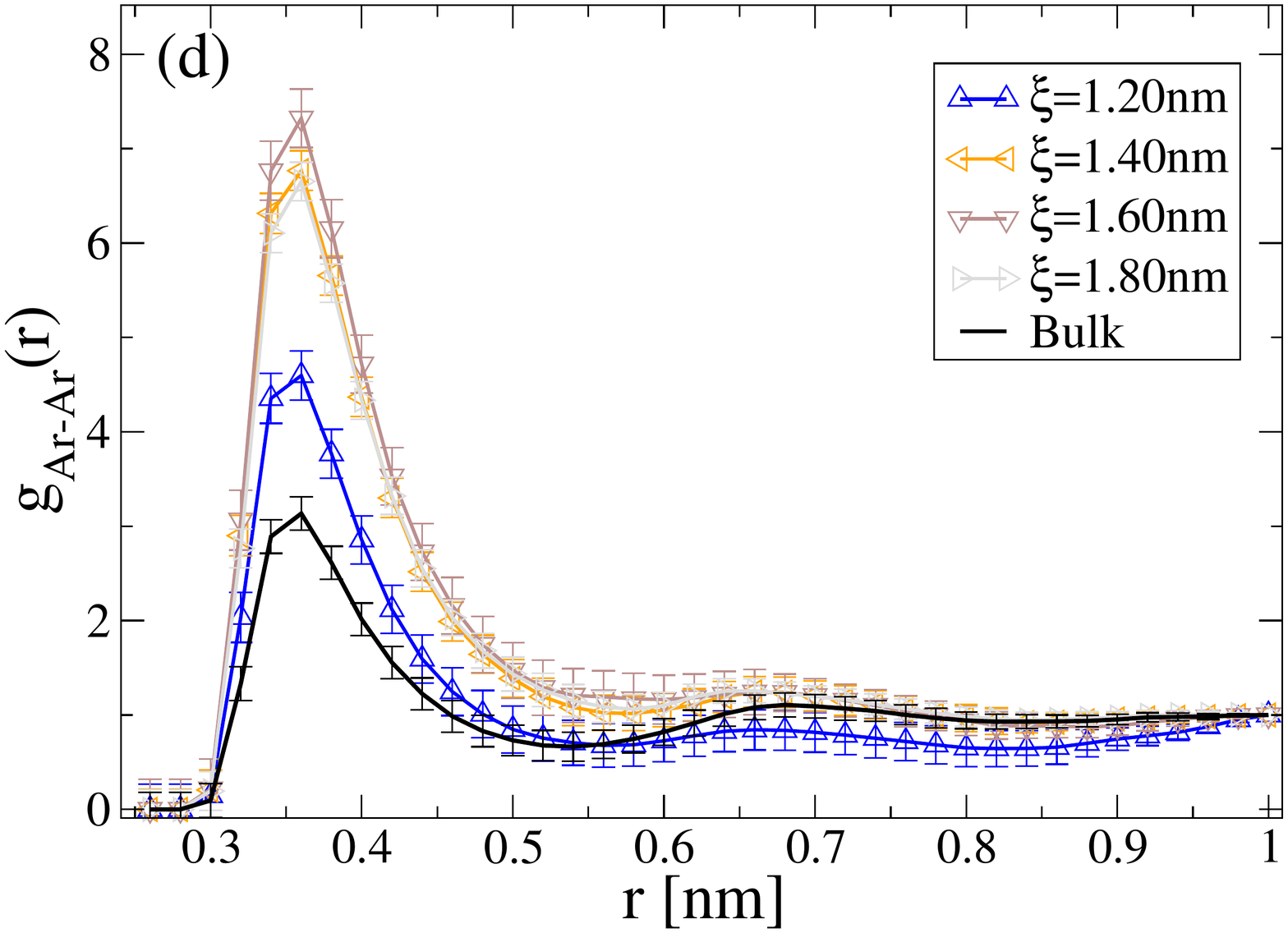}
\end{center}
\caption{(color online) Potential of mean force between argon in
  nanoconfiment for different confinement widths (a)
  $\xi=0.6,0.8,1.0$, and (b) $\xi=1.2,1.4,1.6,1.8$~nm
  respectively. For a comparision, we also show PMF for argon in bulk
  water at $T=300$~K and $P=1$~atm. Argon-argon radial distribution
  function $g_{\rm Ar-Ar}(r)$ calcuated using PMF $w(r)$ for
  confinement widths (c) $\xi=0.6,0.8,1.0$~nm, and (d)
  $\xi=1.2,1.4,1.6,1.8$~nm.}
\label{fig:fig3}
\end{figure}

In Fig.~\ref{fig:fig3}(a) and (b), we show PMF, $w(r)$, as a function
of distance $r$ between argon atoms for all the confinement
widths. For a comparison, we also show the PMF for argons in bulk
water at $P=1$~atm and $T=300$~K. $w(r)$ for bulk system exhibits two
prominent minima, one at $r\approx0.35$~nm and another at $r\approx0.68$~nm
respectively. While the position of the first minimum of $w(r)$
remains unchanged in confinement, the position of the second minimum
decreased to $r\approx0.64$~nm for the smallest $\xi$. Moreover, while
the first minimum becomes deeper monotonically with decreasing
confinement width, the second minimum shows a non-monotonic behavior
with $\xi$ suggesting a non-trivial solvation structure dependence
with confinement width $\xi$.

In Fig.~\ref{fig:fig3}(c) and (d), we show the radial distribution
function between argons as calculated from $g_{\rm Ar-Ar}(r)=e^{-\beta
  w(r)}$ for all the confinement widths studied here. It is clear that
solvation structure of argons in water in nanoconfinement exhibits a
non-trivial dependence on the confinement width of confining
surfaces. Since, this non-trivial behavior of PMF or $g_{Ar-Ar}(r)$
for the second shell makes it harder to interpret the hydrophobicity
directly by looking at them, therefore to quantify hydrophobicity, we
next calculate the second Virial coefficient.

The second Virial coefficient $B_2$ for a mono-component system is
given by
\begin{equation}
B_2 = -\frac{1}{2}\int_{0}^{\infty} (e^{-\beta w(r)}-1) d\Gamma
\end{equation}
where $\beta=\frac{1}{k_BT}$, and $d\Gamma$ is the volume element
corresponding to the separation $r$. $w(r)$ is related to the
solute-solute pair correlation function $h(r)$ as

\begin{equation}
w(r) = - k_BT \ln[h(r)+1]
\end{equation}
and hence
\begin{equation}
B_2 = -\frac{1}{2}\int h(r) d\Gamma
\end{equation}
A positive value of $B_2$ indicates an effective repulsive interaction
between the solutes and hence larger solubility and a negative value
suggest an effective attractive interaction and hence smaller
solubility. The volume element, $d\Gamma=A(r,\xi)dr$, in the planar
confinement depends on both $r$ and $\xi$ and can be derived by first
deriving the average surface area $A(r,\xi)$ corresponding to solute
separation $r$. Let's assume that the surfaces are represented by
two planes; one at $z=0$ and another at $z=\xi$. Assuming that the
first particle can assume any value of $z$ between $0$ and $\xi$, we
can next write the average area, $A(r;\xi)$, traversed by the radial
vector joining two particles as (a detailed derivation of $A(r;\xi)$
is given in the Supplementary Information):
\begin{widetext}
\begin{equation}
A^{*}(r;\xi) =  \left\{ \begin{array}{rl}
4\pi r^2[1-\frac{r}{2\xi}]  &\mbox{ if $r \le \xi/2$} \\
2\pi r\xi (1-(\frac{r}{\xi})^2)+4\pi r^2(1-\frac{\xi}{2r})  &\mbox{ if $\xi/2 \le r \le \xi$} \\
2 \pi r \xi  &\mbox{ if $r \ge \xi$} \end{array} \right.
\end{equation}
\end{widetext}
The above expression suggests that the accessible area to arrange a
pair of particle of fixed separation $r$ decreases with $r$ as
compared to bulk.

In Fig.~\ref{fig:fig4}, we show the second Virial coefficient $B_2$ as
a function of $\xi$. The values of $B_2$ are significantly smaller
compared to the bulk value of $B_2$ ($-17\AA^{3}$) at $P=1$~atm and
$T=300$~K except for $\xi=1.2$~nm, suggesting that the hydrophobicity
of argon increases in nanoconfinement. Moreover, $B_2$ exhibits a
nonmonotonic dependence on $\xi$. $B_2$ first increases with
increasing $\xi$ for $\xi=0.6-1.2$~nm and then it falls off sharply
with a weak dependence on $\xi$ for $\xi=1.4-1.80$~nm.

Since the simulations are done in NVT-ensemble, the pressure for
different values of $\xi$ are different even when the effective
density remains the same. One may argue that the change in
hydrophobicity is just due to pressure increase as the confinement
width decreases. To test this we next calculated vlaues of $B_2$ for
argon for two different pressures $P=1$ and $500$~atm in bulk water. The
pressures were chosen such that they cover the pressure ranges for the
nanoconfined system. In Fig.~\ref{fig:fig5} (a), we show the PMF for
P=1 and 500~atm for the bulk system at $T=300$~K. The second Virial
coefficient $B_2$ for $P=1, 500$~atm are $-17$ and $53$~$A^{o^3}$
respectively. We find that the bulk values of $B_2$ differ greatly
from the nanoconfined values except for $\xi=1.2$~nm, for which the
system shows a comparable value of $B_2$. To this end, it clear that
the magnitude of hydrophobicity in nanoconfinement is drastically
different from the bulk. Moreover, our results indicate subtle changes
in solvation and effective hydrophobicity upon
nanoconfinement. Indeed, small changes in the length scale of
nanoconfining regions may lead to large changes in hydrophobicity.
\section*{From bulk PMF to confined PMF}

\begin{figure}[b]
\begin{center}
\includegraphics[width=12cm]{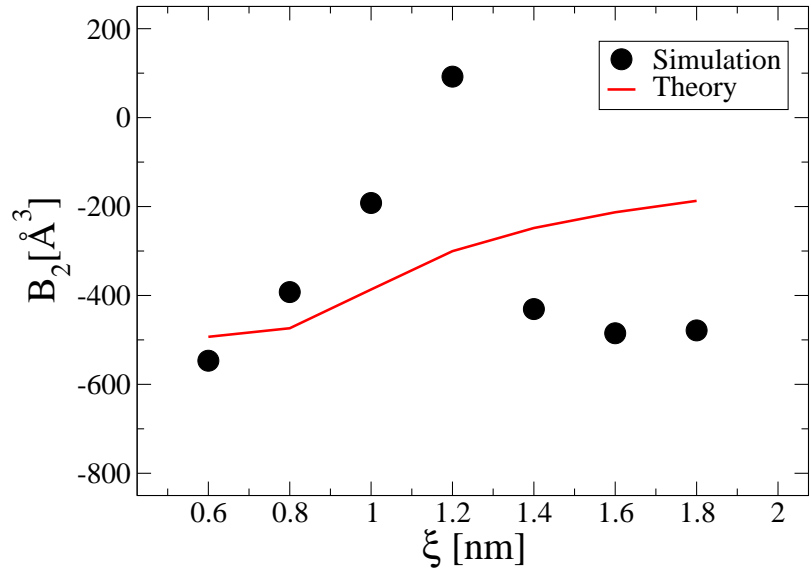}
\end{center}
\caption{(color online) Second Virial coefficient $B_2$ between argons
  as a function confinement width $\xi$. Also shown is the curve obtained
  using entropic consideration in confinement (see Section V).}
\label{fig:fig4}
\end{figure}

\begin{figure}[h]
\begin{center}
\includegraphics[width=10cm]{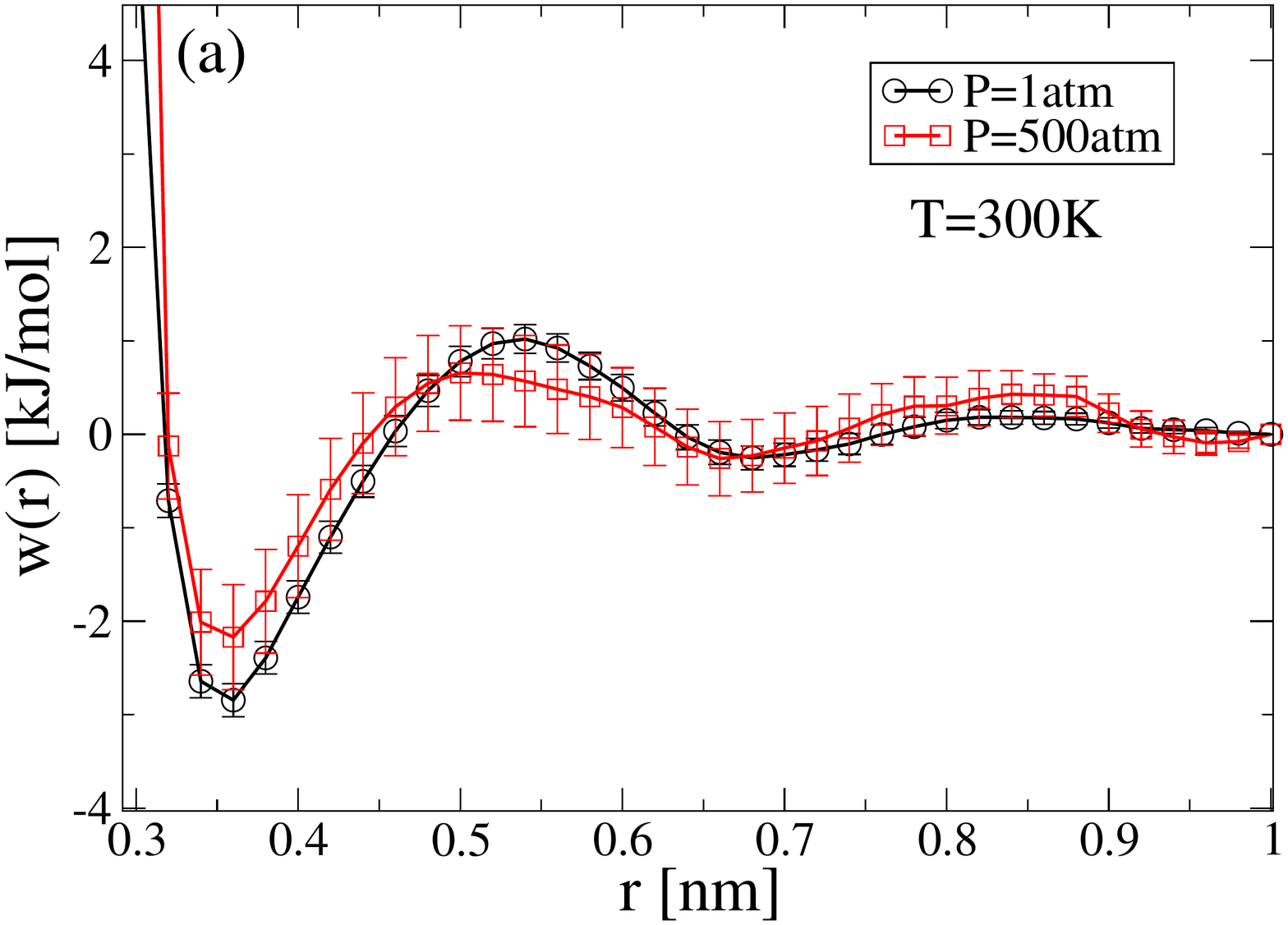}
\includegraphics[width=10cm]{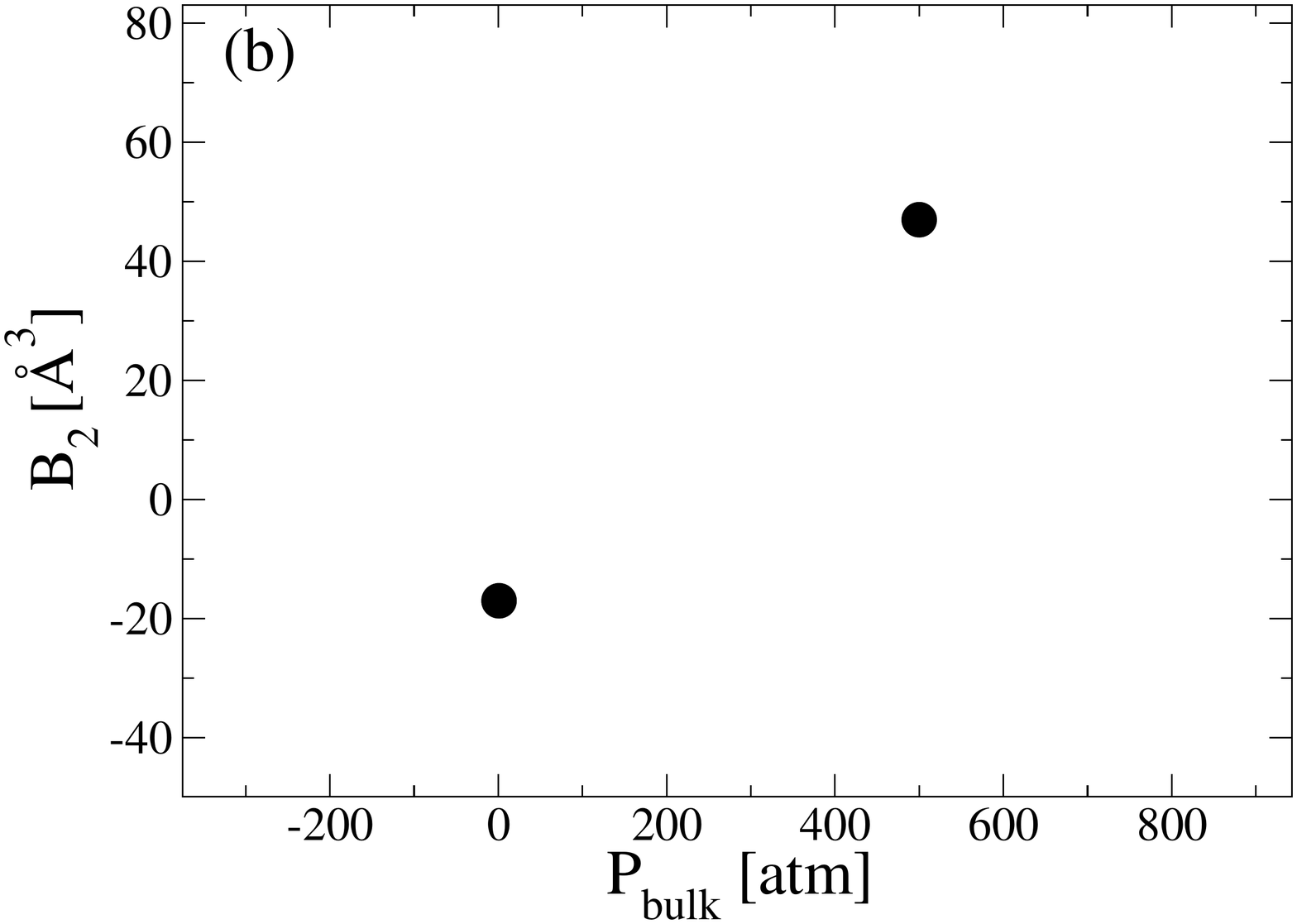}
\end{center}
\caption{(color online) (a) Potential of mean force between argon, $w(r)$,
  for bulk system for two different pressures $P=1,$~and
  $500$~atm and temperature $T=300$~K. (b) Second Virial coefficient
  $B_2$ for two pressures $P=1,$~and $500$~atm and temperature
  $T=300$~K.}
\label{fig:fig5}
\end{figure}

Nanoconfinement results in changes in arranging a solute particle
around a reference particle (see Supplementary Information) --namely a
decrease in the volumetric arrangement, we can define the change in
entropy $\Delta S_1(r;\xi)$ in nanoconfinement over a bulk system as:
\begin{widetext}
\begin{equation}
\Delta S_1(r;\xi) = \left\{ \begin{array}{rl} k_B \ln
  (1-\frac{r}{2\xi}) &\mbox{ if $ 0 \le r \le \xi/2$}
  \\ k_B \ln
     [(1-\frac{\xi}{2r})+\frac{\xi}{2r}(1-(\frac{r}{\xi})^2)] &\mbox{
       if $\xi/2 \le r \le \xi$} \\  k_B \ln
     (\frac{\xi}{2r}) &\mbox{ if $ \xi \le r \le r_0$} \end{array} \right.
\end{equation}
\end{widetext}
where $r_0$ is the separation between solutes such that ${\rm
  lim}_{r->r_0}w(r)=0$. Note that $r_0$ is the distance beyond which
the system loses two-point correlations. In our case, we assume
$r_0=1$~$nm$. The validity of above expression ranges for the values
of $r$ over which the solute particles are correlated beyond which
the two point entropy would be zero for both bulk and confined
system. Since the entropy decreases as $r$ is increased, we argue that
the system with solute nanoconfinement will try to increase the
configurations pertaining to larger entropy in order to minimize free
energy. Hence, we can assume an additional thermodynamic driving
potential $-T\Delta S_1(r;\xi)$ acting on a pair of particles. This
entropic penalty will make small distances between solutes
more favorable as they correspond to higher entropy. Taking this into
account, we can write the modified PMF in nanoconfinement $w^*(r;\xi)$
as
\begin{equation}
w^*(r;\xi) = w_{\rm bulk}(r) - T\Delta S_1(r) 
\label{eq:modwr}
\end{equation}
The $w^{*}(r;\xi)$ is then normalized such that $w^*(r=1nm;\xi)=0$. In
Fig.~\ref{fig:fig7}, we compare PMF computated from simulations and
PMF from Eq.~\ref{eq:modwr} for confinement widths $\xi=0.60$~nm,
$\xi=0.80$~nm, $\xi=1.60$~nm, and, $\xi=1.80$~nm. We find that the PMF
predicted by theory does reasonably well for smaller $\xi$ but it
deviates from the PMF calculated from simulations for larger
$\xi$. The deviation of theory could presumably results from absence
of enthalpic consideration. We show the values of $B_2$ calculated
from $w^*(r;\xi)$ for different values of confinement widths in
Fig.~\ref{fig:fig4} along with the computed values of $B_2$ from
simulations.

We find that simple argument of entropy decrease of configurations
with different separation of a pair of solute atoms in confinement
works well with planar geometry studied here. Indeed, a similar
argument can be made for different geometries of confinement such as
cylindrical nanoconfinement.

\section*{Solvation Structure}
Next, we investigate the solvation structure in order to find a clue
to the anomalous hydrophobicity behavior for $\xi=1.2$~nm by
investigating the radial distribution of water and argon. For water,
we calcualte the oxygen-oxygen lateral radial distribution function,
$g_{\rm oo}(r_{xy})$ as a function their lateral distance $r_{\rm
  xy}$. In Fig.~\ref{fig:fig8} (a), we show $g_{\rm oo}(r_{\rm xy})$
for all the confinement widths $\xi$. We find that, water tends to
order laterally with the decrease of confinement width as suggested by
increased first and second peaks in $g_{\rm oo}(r_{\rm xy})$ upon
decrease of $\xi$. In Fig.~\ref{fig:fig8}(b), we show the radial
distribution function between oxygen of water and argon, $g_{O-Ar}(r)$
calculated from long simulations of water-argon system ($4$~ns for
each confinement width). While the structure of the first solvation
shell ($r\approx 0.36$~nm) does not show any appreciable change with
$\xi$ except the decreasing value of the first peak, the second shell
becomes much wider and the corresponding peak moves to slightly larger
values of $r$. Combining the results on PMF between argon and
solvation structure, it seems that non-monotonic dependence of second
shell on $\xi$ may explain the increased value of $B_2$ for
$\xi=1.2$~nm and will require further investigation.

\section{Summary and Discussion}
We have studied molecular scale hydrophobicity of a small apolar
solute in nanoconfinement by calculating the second Virial
coefficients from the potential of mean force between apolar solutes
in water confined between hydrophobic surfaces with different
confinement widths. We find that: (i) hydrophobicity of apolar solutes
usually increases as the confinement width decreases at constant
transverse pressure, (ii) hydrophobicity exhibits an
anomalous region, where the hydrophobicity becomes similar to the
values in bulk water. Furthermore, we find that this anomalous region
of confinement widths correspond to changes in second and third
solvation shell. We develop a simple entropic theory to find effective
potential of mean force in nanoconfinement. The predicted PMF from
theory works reasonably well for smaller confinement widths and allows
to predict the effective hydrophobicity in nanoconfinement when the
potential of mean force in bulk is known.

Hydrophobicity plays important roles in many physical systems
including the folding of polypeptides. We hypothesize that large
changes in hydrophobicity with very small changes in length scale of
nanoconfining regions could be very important in maintaining the rate
of folding in biological systems such as chaperone proteins in a
chemically non-specific way. A very good example of this is
GroEL-GroES complex in bacteria, which is induced upon temperature
shocks~\cite{cage}. To counteract the increase of temperature and hence
unfolding of polypeptides, nature has evolved a complex machinery. An
unfolded polypeptide is directed into a chaperon complex such as
GroEL-GroES and stays inside the cage formed by the protein complex
where it folds much faster than it would in the bulk
region~\cite{cage}. We argue based on our results that when and
unfolded or partially folded polypeptide enters this complex, a large
part of the folding may occur in the neck-region before it reaches the
cage. Moreover, we hypothesize that the modulating diameter of this
chaperon complex would naturally be helpful in modulating the rate of
folding in a chemically non-specific way.

In summary, our results are important in understanding nanoconfinement
induced stability of apolar polymers, solubility of gases and may help
design better systems for enhanced oil recovery. While our work has
explored the planar hydrophobic confinement in details, the geometry
and charge distribution of the confining surfaces may affect the
hydrophobicity. Furthermore, effect of solute size consideration is
also a determining factor for hydrophobicity which we will explore in
future work.

\section*{Acknowledgment}
Authors would like to thank Harpreet Kaur, Khanh Nguyen, William Lin
Oliver, and Paul Thibado for fruitful discussion and University of
Arkansas High Performance Computing Center for providing computional
time.

\clearpage

\begin{figure}
\begin{center}
\includegraphics[width=12cm]{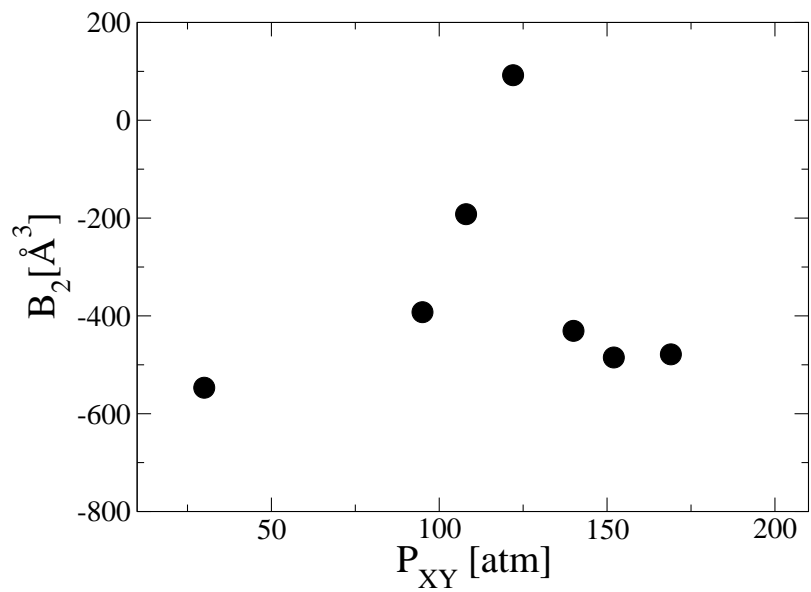}
\end{center}
\caption{Second Virial coefficient $B_2$ as a function of lateral pressure $P_{\rm xy}$.}
\label{fig:fig6}
\end{figure}
\begin{figure}
\begin{center}
\includegraphics[width=8cm]{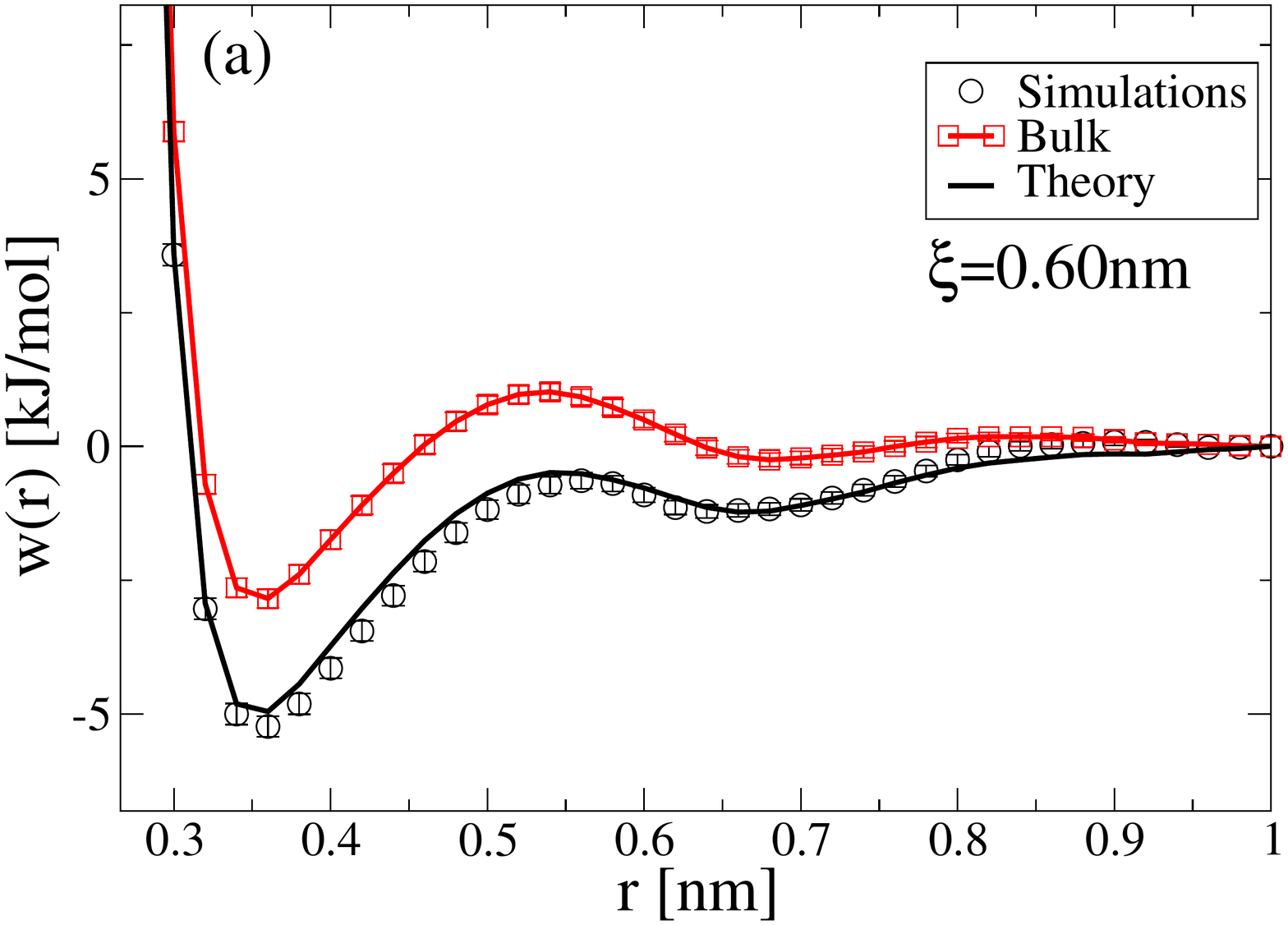}
\includegraphics[width=8cm]{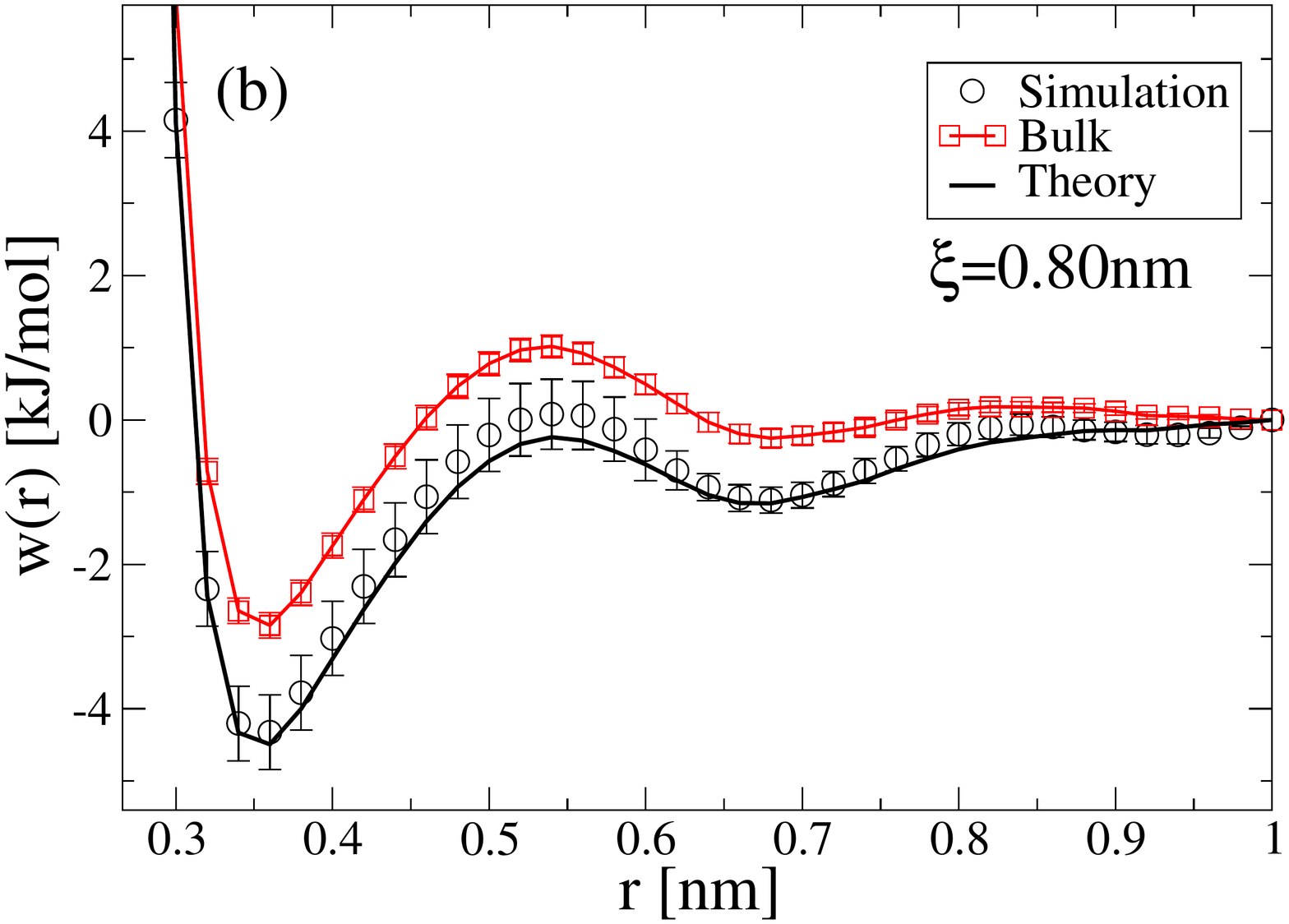}
\includegraphics[width=8cm]{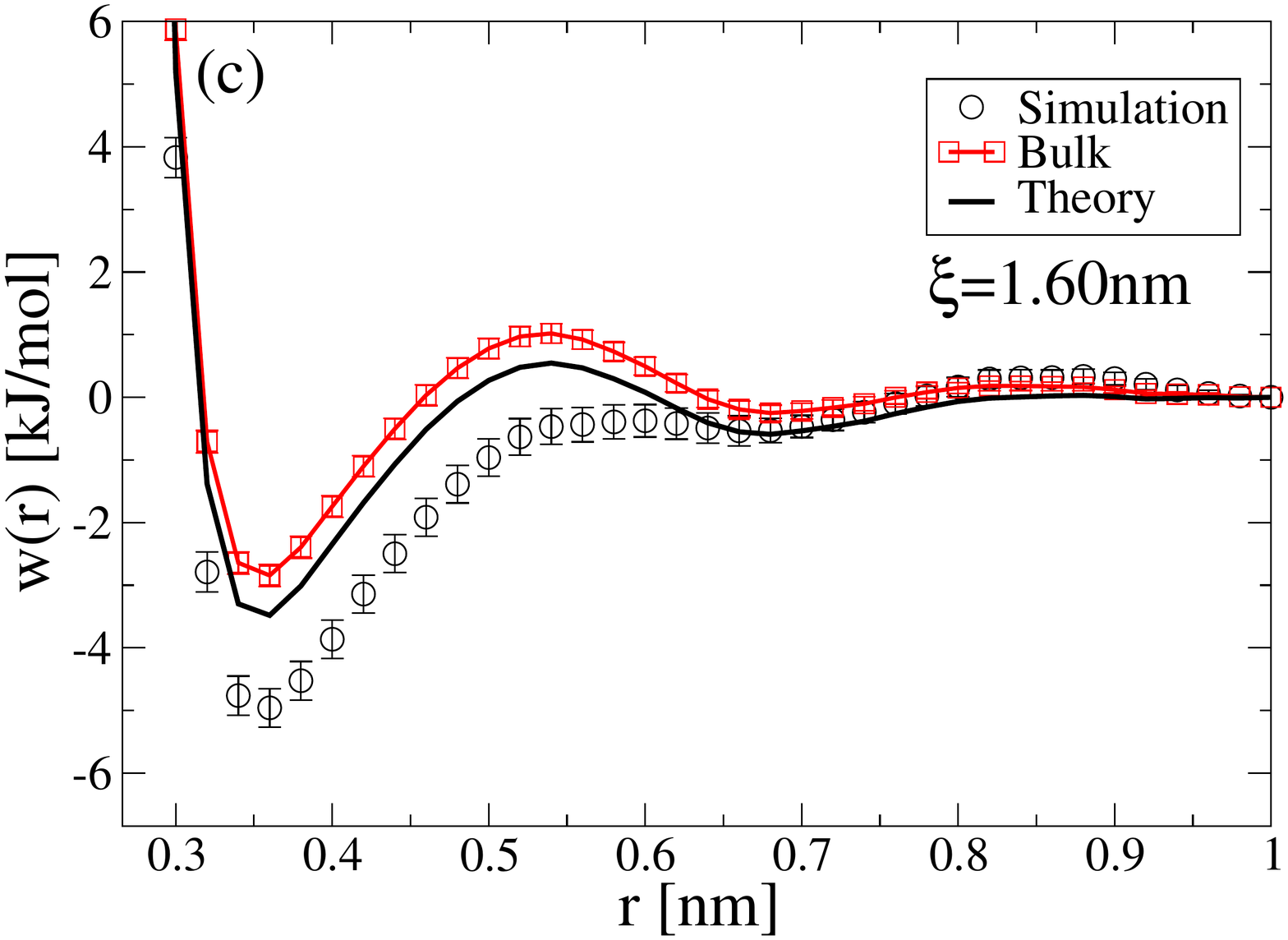}
\includegraphics[width=8cm]{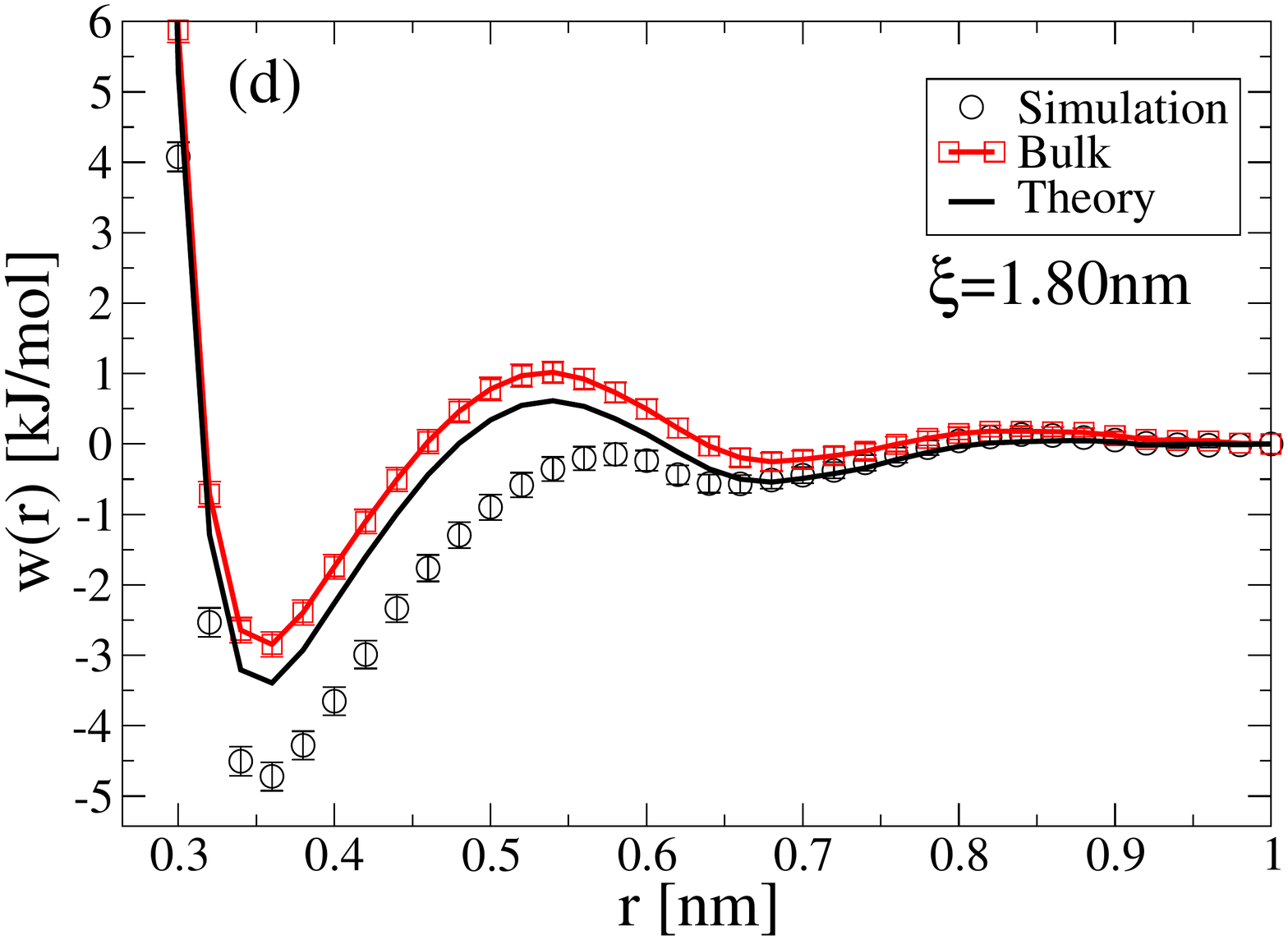}
\end{center}
\caption{(color online) Comparision of potential of mean force calculated from
  simulation with theory for confinement widths (a) $\xi=0.60$~nm, (b)
  $\xi=0.80$~nm, (c) $\xi=1.60$~nm, and (d) $\xi=1.80$~nm. While the
  PMF predicted from theory does reasonably well for smaller $\xi$, it
  deviates from the PMF calculated from simulations for larger $\xi$.}
\label{fig:fig7}
\end{figure}
\begin{figure}
\begin{center}
\includegraphics[width=8cm]{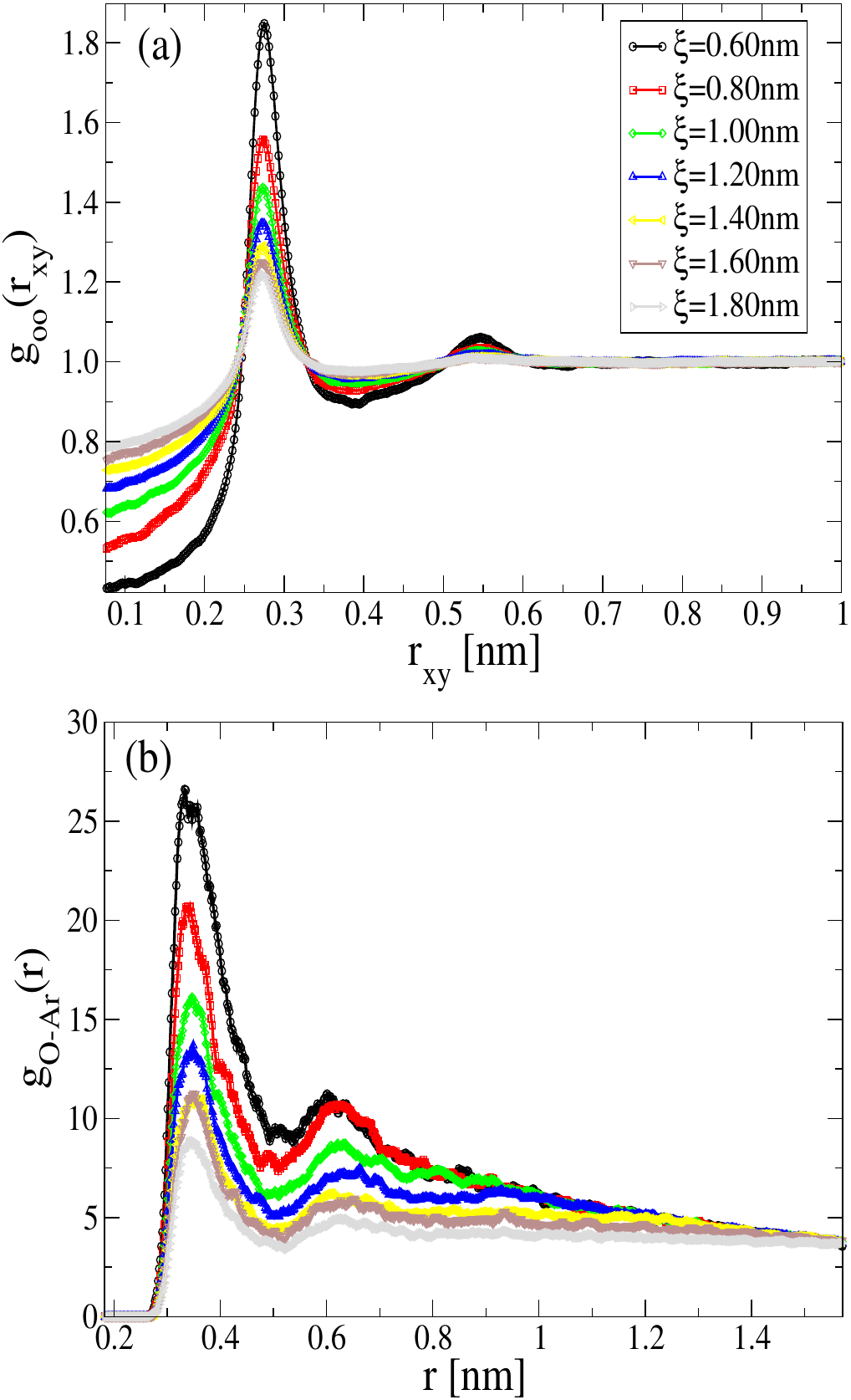}
\end{center}
\caption{(a) Lateral oxygen-oxygen radial distribution function,
  $g_{\rm oo}(r_{\rm xy})$ of water for various confinement width. (b)
  Oxygen-Argon radial distribution function, $g_{\rm O-Ar}(r)$, for
  all the confinement widths studied here.}
\label{fig:fig8}
\end{figure}

\clearpage

\appendix

\renewcommand{\thefigure}{S\arabic{figure}}

\setcounter{figure}{0}

\noindent {\bf \large Supplementary Information:}

\bigskip
\bigskip

\noindent {\bf Effective surface area traced by a particle at a distance $r$ from another particle}

\bigskip
\bigskip

\noindent Let's assume that two infinite planar surfaces are located
at $z=0$ and $z=\xi$ respectively. We are interested in the average
surface area traced by a particle at a fixed distance $r$ from the
reference particle. It is easy to see that the average surface area
for a fixed $r$ depends on both $r$ and $\xi$. When the reference
particle is closer to the surface the area is smaller compared to when
the reference particle is sitting close to the center of the surfaces.
\begin{figure}[hb]
\begin{center}
\includegraphics[width=12cm]{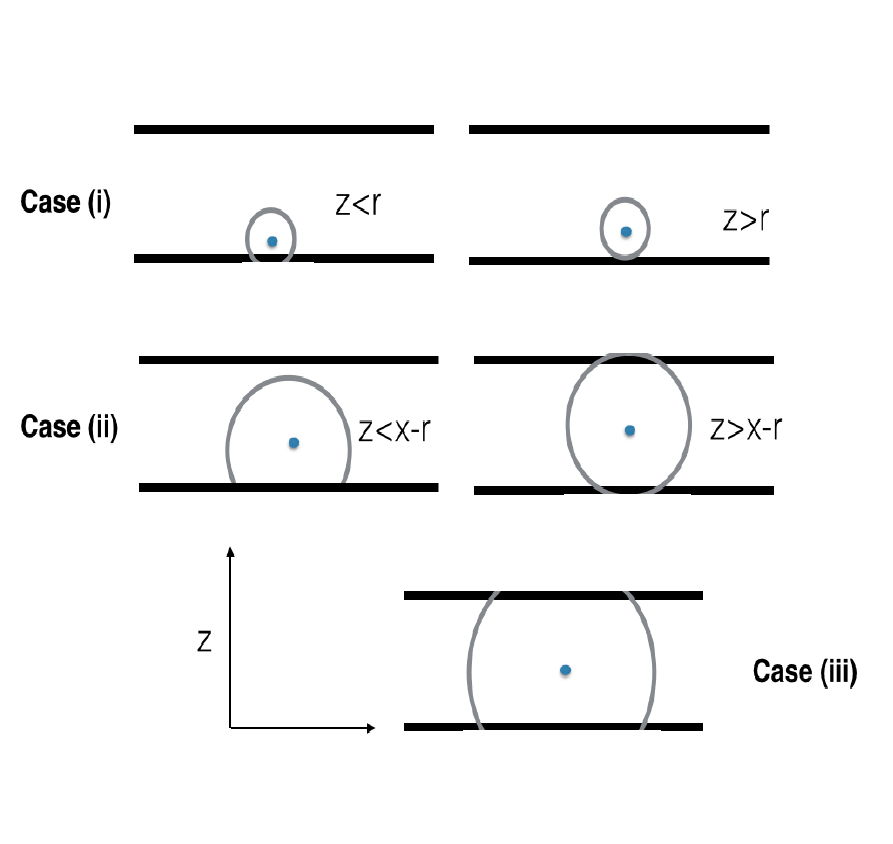}
\end{center}
\caption{Schematics of the regions of $r$ discussed above for a confinement width $\xi$.}
\label{fig:figSchema}
\end{figure}
Three different regimes of $r$ -- (i)  $0 \le r\le \xi/2$, (ii)  $\xi/2 \le r\le \xi$, and (iii) $r \ge \xi$ can be noted. Furthermore, from the symmetry of the system, it is sufficient to calculate the average area traced by the second particle for different positions of the reference particle only along the $z$ direction. Let's assume that the reference particle is sitting at a distance $z$ from the surface at $z=0$. In the following, we will derive the expression for the average surface area traced by a second particle placed at a fixed distance $r$ from the reference particle.  
\bigskip
\noindent  {{\bf Case (i)} : $0 \le r\le \xi/2$}
\bigskip

The average area $A(r;\xi)$ is given by 
\begin{equation*}
A(r;\xi) = \frac{1}{\xi} \int_{0}^{\xi}A(r,z;\xi) dz
\end{equation*}
where $A(r,z;\xi)$ is the surface area traced by the second particle when the reference particle is located at a distance $z$ along the $z$-direction. For this case, we can split $A(r,z;\xi)$ into two terms, (a) when $z \ge r$  or $\xi-z \ge r$ and (b) $ z < r $ or $\xi-z < r$ (see Fig.~\ref{fig:figSchema}). Hence, the total area $A(r,z;\xi)$ can be written as 
\begin{equation*}
A(r,z;\xi) =  4 \pi r^2 (\xi-2r) + 2\int_{0}^{r} 2\pi (r+z) dz = 4 \pi r^2 \xi (1-\frac{r}{2\xi})
\end{equation*}
And hence, the average area $A(r,\xi)$ can be written as
\begin{equation*}
A(r;\xi) = 4 \pi r^2 (1-\frac{r}{2\xi})
\end{equation*}
\bigskip

\noindent  {{\bf Case (ii)} : $\xi/2 \le r\le \xi$}

\bigskip
In this case, we can show that the total area $A(r,\xi)$ can be written as a combination of two terms when (a) $z < \xi-r$, and (b) $\xi-r < z \le \xi/2$ (see Fig.~\ref{fig:figSchema}), leading to 
\begin{equation*}
A(r,z,\xi) = 2 [\int_{0}^{\xi-r} 2 \pi r (r+z) dz + 2\pi r \xi (r-\xi/2)] = 2 \pi r \xi (r-\frac{\xi}{2}) + \pi r \xi^2 (1-\frac{r^2}{\xi ^2})
\end{equation*}
And hence the average area, $A(r;\xi)$ is given by 
\begin{equation*}
A(r;\xi) = 4 \pi r^2 (1-\frac{\xi}{2 r}) + 2\pi r \xi (1-\frac{r^2}{\xi^2})
\end{equation*}

\noindent  {{\bf Case (iii)} : $ r \ge \xi$}

\begin{figure}
\begin{center}
\includegraphics[width=10cm]{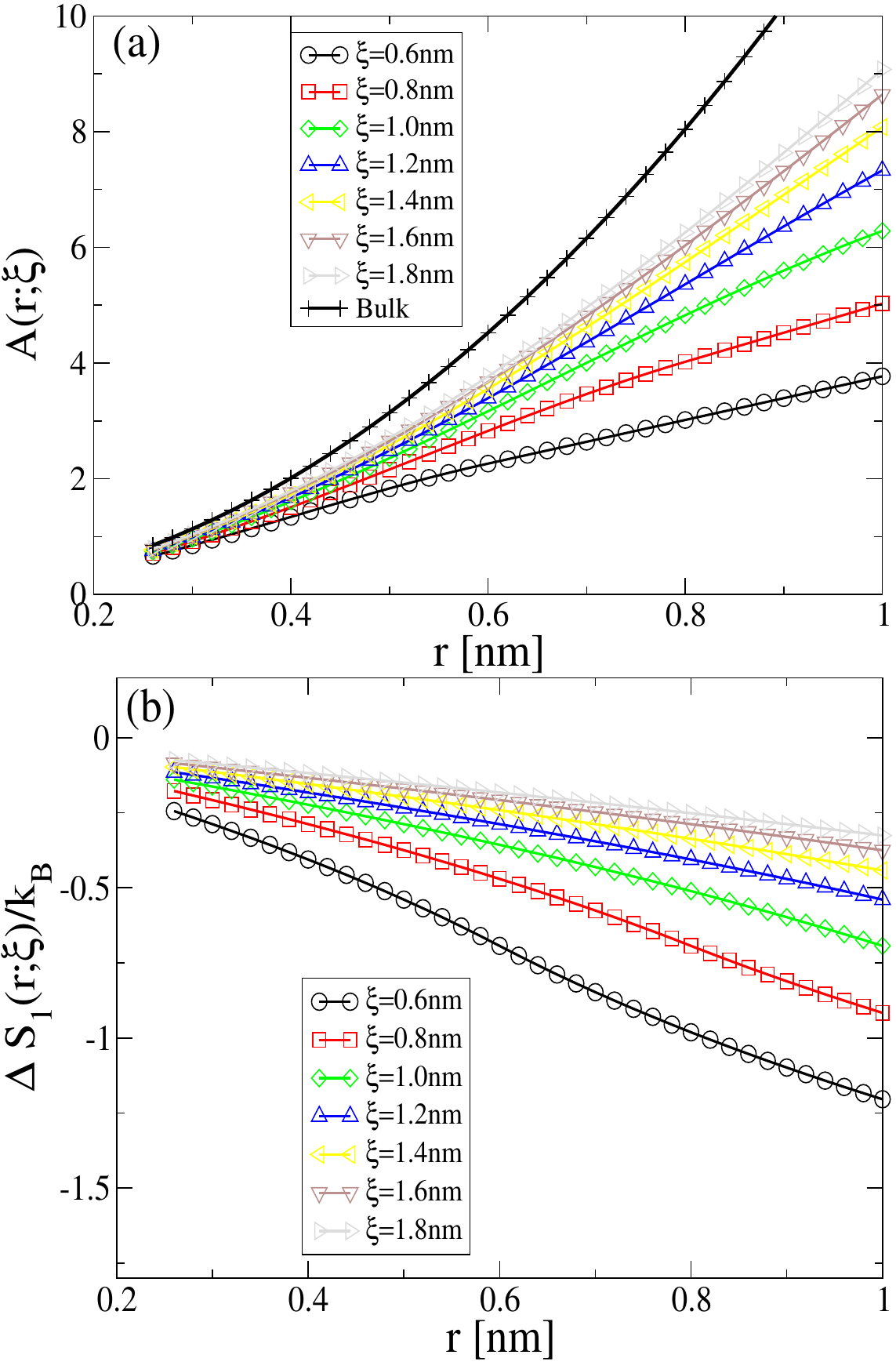}
\end{center}
\caption{(a) Effective average area traced by a particle for a fixed distance $r$ from a reference particle, and (b) Entropy change $\Delta S_1(r;\xi)$ as a function of $r$ for different confinement widths.}
\label{fig:figEntropy}
\end{figure}

\bigskip
In this case the total area $A(r,z;\xi)$ is independent of the location $z$ of reference particle  and is  $2 \pi r \xi$ and hence the average area $A(r;\xi)$ is given by (see Fig.~\ref{fig:figSchema})
\begin{equation*}
A(r;\xi) = 2 \pi r \xi
\end{equation*}
In Figure~\ref{fig:figEntropy} (a), we show the effective average area $A(r;\xi)$ as a function of $r$ for all the confinement widths studied here. Also, in Figure~\ref{fig:figEntropy}(b), we show entropy change $\Delta S_1(r,\xi)$ as a function of $r$. Entropy change, $\Delta S_1(r;\xi)$, over the bulk system is defined as:
\begin{equation}
\Delta S_1(r;\xi) = k_B {\rm log}(\frac{A(r,\xi)}{4\pi r^2})
\end{equation}
where $k_B$ is the Boltzmann constant. $\Delta S_1(r;\xi)$ represents the change in entropy of volumetric arrangement of a two-particle system  in confinement over the bulk situation. As expected, the entropy change for small $r$ in confinement is smaller compared to large $r$ which gets increasingly larger as the confinement width is reduced.

\bigskip

\end{document}